\documentclass{ws-mpla}
\usepackage[super]{cite}
\usepackage{graphicx}
\begin{document}

\markboth{Authors' Names}
{Instructions for Typing Manuscripts (Paper's Title)}

\catchline{}{}{}{}{}

\title{A PROPOSAL TO AVOID THE AMBIGUITY IN THE IDENTIFICATION OF THE SCALE ENERGY  PARAMETER $\tilde{Q}^{2}$ OF THE PDFs IN THE $Z$-PRODUCTION IN $ep$-DIS\footnote{\bf Dedicated to the memory of Prof. Dr. Manfred  B\"ohm (27.11.1940 - 5.7.2013). }}

\author{M. G\'OMEZ-BOCK}

\address{DAFM, Universidad de las Am\'ericas Puebla, Ex. Hda. Sta. Catarina M\'artir s/n,\\
San Andr\'es Cholula, Puebla. M\'exico.\footnote{melina.gomez@udlap.mx}}

\author{W. GONZALEZ and A. ROSADO}

\address{Instituto de F\'{\i}sica, Benem\'erita Universidad Aut\'onoma de Puebla,\\Apartado Postal J-48, C.P. 72570 Puebla, Pue., M\'exico.
 \footnote{wendyg@ifuap.buap.mx, rosado@ifuap.buap.mx}}

\author{L. T. L\'OPEZ-LOZANO}
\address{Universidad Aut\'onoma del Estado de Hidalgo, Carr. Pachuca-Tulancingo \\Km. 4.5, C.P. 42184, Pachuca, Hidalgo, M\'exico.\footnote{lao-lopez@uaeh.edu.mx}}

\maketitle

\pub{Received (Day Month Year)}{Revised (Day Month Year)}

\begin{abstract}
We discuss $Z$-production in the process $e + p \to e + Z + X$ in the context of the Standard Model using the Parton Model. In contrast to the  inelastic $ep$-scattering ($e + p \to e + X$), where the choice of $\tilde{Q}^{2} $ is unambiguous, in this case is not clear since the momentum transfer square, at the quark level, depends on the reaction mechanism. Our aim in this work is to show that the total cross section depends strongly on the prescription used for the scale energy parameter $\tilde{Q}^{2}$ and on a different option that we have considered for making the convolution of the parton distribution functions and the amplitude of the quark processes. We present results for the total cross section as a function of the total energy $\sqrt{s}$ of the system $ep$, in the range $300 \leq \sqrt{s} \leq  1,300$ GeV. We find differences of up to 25\% in the rates of the total cross section for the different prescriptions that we have taken. For $\sqrt{s} \approx 1,300$ GeV, which is the expected maximal total energy of the system $ep$ to be reached at the LHeC and taking an integrated luminosity of 0.1 $ab^{-1}/year$, such a difference (25\%) corresponds  to $\approx 10^4$ produced $Z$ bosons. Finally, we propose to introduce a slight modification in the form that we make the convolution of the PDFs and the amplitudes of the subprocesses involved in the process in order to get a simple and practical prescription for the calculation without ambiguities of the $Z$-production through the deep inelastic $ep$ scattering, by using the Parton  Model.
\keywords{Z production; Deep Inelastic Scattering; Scale energy parameter}
\end{abstract}

\ccode{12.15.Ji, 12.15.Mm, 13.60.-r, 13.60.Hb, 14.70.Hp}

\section{Introduction}   
LHeC at CERN will provide us with the possibility to observe $ep$ collisions with a maximal energy $E^{max} = 60$ GeV of the electron and $E^{max}_p = 7$ TeV of the proton \cite{partdata}, this means that the maximal total energy of the $ep$ system will be $\sqrt{s} \approx$ 1,300 GeV. The LHeC is a potentially rich source of $Z$ bosons produced by electron-proton collisions. The LHeC will make possible to increase de number of produced $Z$ bosons {\it via} ep collisions up to three orders of magnitude, with respect to HERA. This means that the LHeC will be an excellent collider for the search of physics beyond the Standard Model \cite{stanmod}. However, this can be done only if the results of the Standard Model are well known in all details and without ambiguities. Therefore, we discuss in this work the possibility of finding a prescription to calculate $Z$-production unambiguously through the deep inelastic process $e + p \to e + Z + X$ using the Standard Model and by making use of the Parton Model \cite{partmod1}. We perform our numerical calculations using the QCD improved Parton Model \cite{partmod2,partmod3}. According to the Parton Model, the final step in the evaluation of $d\sigma^{ep}$ consists in putting together the parton cross sections $d\sigma^{eq}$ and the parton distribution functions (PDFs) $ f_q (x^{\prime}, \tilde{Q}^{2})$ and to make a sum over all the possible partons. In the PDFs $x^{\prime}$ is the fraction of momentum that the incoming quark carries of the colliding proton and the parameter $\tilde{Q}^{2}$ stands for a scale energy, usually identified as the momentum transfer square in the collisions. At the lowest order in $\alpha$, in contrast to deep inelastic $ep$ scattering ($e + p \to e + X$) where the choice of $\tilde{Q}^{2} $ is unambiguous, in the case of $Z$-production this is not clear since the momentum transfer square depends on the reaction mechanism, in other words, whether the $Z$ is emitted at the lepton or at the quark line. Our aim in this note is to show the strong dependence of the total cross section on the prescription used for the scale parameter $\tilde{Q}^{2}$ and the way that one makes the convolution of the PDFs with the amplitude of the quark processes. At the lowest order in $\alpha$, only two  types of reaction mechanisms will contribute: $Z$ production at the lepton line and at the quark line at the parton level (See Fig.~\ref{fig1}). Diagrams containing the exchange of Higgs bosons will be neglected because of the smallness of the Higgs-fermion coupling involved in the mentioned process. We present results for the total cross section as a function of the total energy $\sqrt{s}$ of the system $ep$, in the range $300 \leq \sqrt{s} \leq  1,300$ GeV. Taking $\sqrt{s} \approx 1,300$ GeV, we find a up to difference of 25\% in the rates of the total cross section for the different prescriptions that we have adopted. This difference (25\%) corresponds to $\sim 10^4$ produced $Z$ bosons, at the planned LHeC by taking an integrated luminosity of 0.1 $ab^{-1}/year$ \cite{partdata}. We perform our numerical calculations using the parton distribution functions PDFs reported by Pumplin {\it et al} \cite{Pumplin,Stump}, we use the CTEQ PDFs provided in a $n_f=5$ active flavors scheme.
Finally, we propose to introduce a slight modification in the form that we make the convolution of the PDFs and the amplitudes of the subprocesses involved in the process in order to get a simple and practical prescription for the calculation without ambiguities of the $Z$-production through the deep inelastic $ep$ scattering, by using the Parton  Model.

\section{Kinematics and Formulae}

In Ref.[8], it has been presented the kinematics and the formulae required in order to calculate the cross section of the production of a $Z$ boson through the inclusive process
\begin{equation}
e + p \to e + Z + X
\end{equation}
We shall denote the four-momenta of these particles by $p,\ P_p,\ p^\prime$ and $k$, respectively. $X$ stands for $anything$.

\noindent As usual the following invariants are defined \cite{Byckling}:
\begin{equation}
\begin{array}{lll}
s   & = &  (p + P_p)^2\\
Q^2 & = & - (p - p^\prime)^2\\
\nu & = & P_p (p - p^\prime)\\
s^\prime & = & (p + P_p - k)^2\\
Q^{\prime 2} & = & - (p - p^\prime - k)^2 \\
{\nu}^{\prime} & = & P_p (p - p^\prime - k)\\
\end{array}
\end{equation}
and the dimensionless variables:
$$x=\displaystyle\frac{Q^2}{2\nu},\hspace{.5cm}
y=\displaystyle\frac{2\nu}{s},\hspace{.5cm}
\tau=\displaystyle\frac{s'}{s},\hspace{.5cm}
x'=\displaystyle\frac{Q'{}^2}{2\nu'},\hspace{.5cm}
y'=\displaystyle\frac{2\nu'}{s}.\hspace{.5cm}
$$
The physical region of these kinematical variables has been discussed in detail in Ref.[8].

\begin{figure}[ht]
\centerline{\includegraphics[width=6.0in]{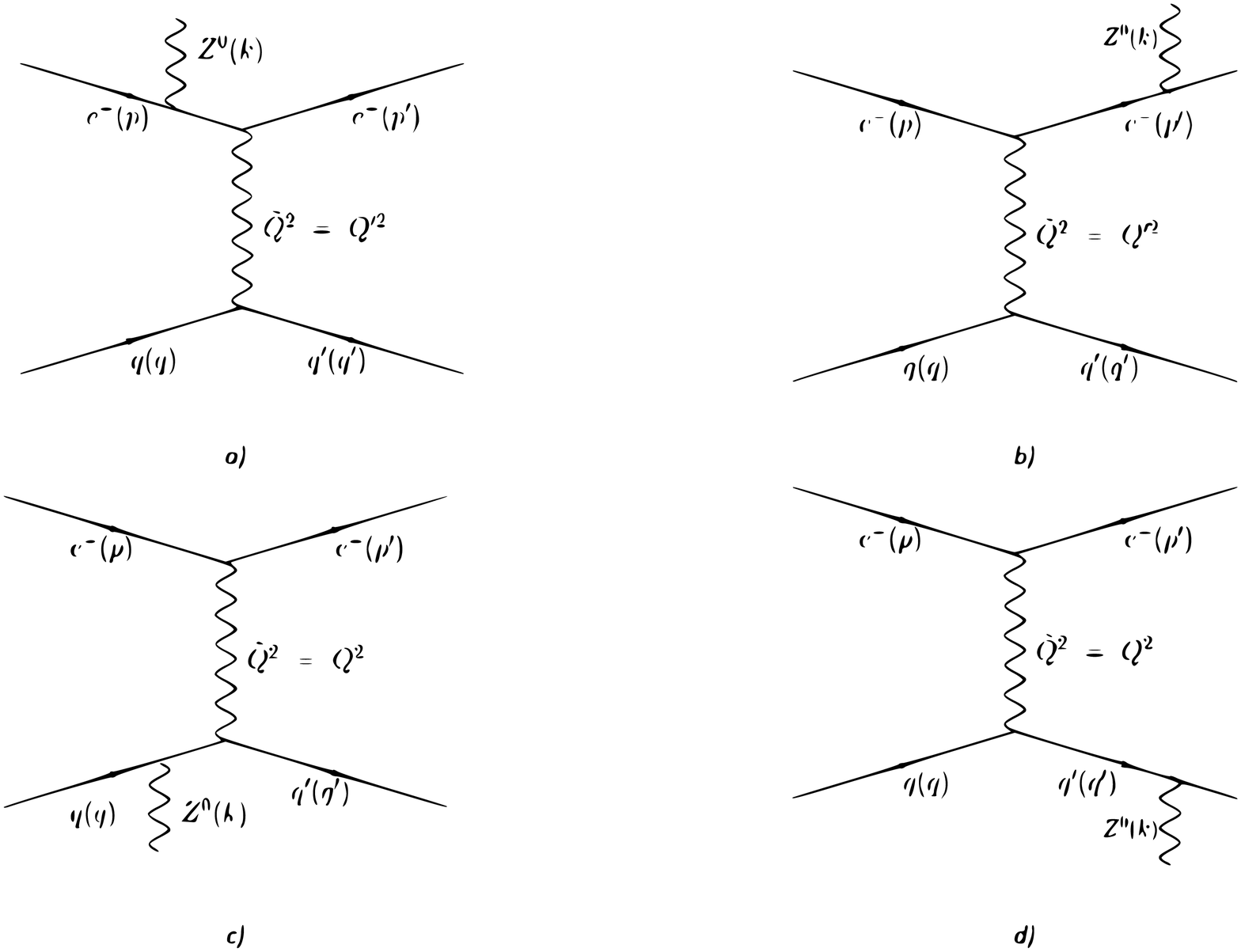}}
\vspace*{8pt}
\caption{Feynman diagrams which contribute at the lowest order in $\alpha$, to $Z$-production through the process $e + p \to e + Z + X$, at the quark level, emitted from the initial \text(a) and final \text(b) electron, the nitial \text(c) and final \text(d) quark.\protect\label{fig1}}
\end{figure}

\noindent The quark cross section is obtained from the invariant matrix element ${\cal M}(e q \to e q Z)$:
\begin{equation}
d\sigma^{eq}_{tot}= \frac{2(2\pi^{-5})}{\hat{s}}
              \frac{1}{4} \mid {\cal M}^{eq}_{tot} \mid^{2} d\Gamma_3
\end{equation}
The 3-particle phase space $d\Gamma_3$ can be expressed with help of the dimensionless set of variables
\begin{equation}
\begin{array}{lll}
d\Gamma_{3} & = &  \displaystyle\frac{\pi s^{3}}{32}\ y\ \displaystyle\frac{dx\ dy\ dy^\prime\
d\tau}{\sqrt{-\Delta_4 \  (p,\ P_p,\ p^\prime,\ k)}}
\end{array}
\end{equation}
with $\Delta_4 (p, P_p, p^{\prime}, k)$ as Jacobi determinant. The Feynman diagrams which contribute at the lowest order in $\alpha$ to ${\cal M}^{eq}_{tot}$ are depicted in Fig.~\ref{fig1}. As we have already  said, the $Z$ boson can be produced from the electron line (Fig.~\ref{fig1}(a),(b)) and the quark line (Fig.~\ref{fig1}(c),(d)), then we write:
\begin{equation}
{\cal M}^{eq}_{tot} \ =\ {\cal M}^{eq}_{l} \ +\ {\cal M}^{eq}_{q}
\end{equation}
Explicit expressions for the quantities needed for the calculation of $\mid {\cal M}^{eq}_{tot} \mid^2$ are presented in Ref.[8], also the sum over the polarizations of the produced boson is performed there.

\noindent The final step in the evaluation of $d\sigma^{ep}$ consists now in putting together the parton cross sections $d\sigma^{eq}$ and the parton distribution functions $ f_q (x^{\prime}, \tilde{Q}^{2})$. In contrast to deep inelastic $ep$ scattering the choice of $\tilde{Q}^{2} $ is not unambiguous in the case of the $Z$-production, since the momentum transfer square to the proton depends on the reaction mechanism (in other words, whether the boson is emitted at the lepton or at the quark line). In order to make clear how strong depend the cross section rates on the choice of the scale $\tilde{Q}^{2}$, we calculate in this work with the following simple prescriptions.

\noindent{Prescription A}
\begin{equation}\label{EqPresA}
\begin{array}{lll}
d\sigma^{ep} & = & {\displaystyle\sum_q} ( \int d x^{\prime} f_q (x^{\prime},M^2_Z)\cdot d\sigma^{eq}_{lepton}\\
&& + \int d x^{\prime} f_q (x^{\prime},M^2_Z) \cdot d\sigma^{eq}_{quark}\\
&& + \int d x^{\prime} f_q (x^{\prime},M^2_Z) \cdot d\sigma^{eq}_{inter} )
\end{array}
\end{equation}
\noindent{Prescription B}
\begin{equation}\label{EqPresB}
\begin{array}{lll}
d\sigma^{ep} & = & {\displaystyle\sum_q} ( \int d x^{\prime} f_q (x^{\prime}, \ Q^2)\cdot
d\sigma^{eq}_{lepton}\\
&& + \int d x^{\prime} f_q (x^{\prime}, \ Q^ 2) \cdot d\sigma^{eq}_{quark}\\
&& + \int d x^{\prime} f_q (x^{\prime}, \ Q^2) \cdot d\sigma^{eq}_{inter} )
\end{array}
\end{equation}
\noindent{Prescription C}
\begin{equation}\label{EqPresC}
\begin{array}{lll}
d\sigma^{ep} & = & {\displaystyle\sum_q} ( \int d x^{\prime} f_q (x^{\prime}, \ Q^{\prime 2}) \cdot d\sigma^{eq}_{lepton}\\
&& + \int d x^{\prime} f_q (x^{\prime}, \ Q^{\prime 2}) \cdot d\sigma^{eq}_{quark}\\
&& + \int d x^{\prime} f_q (x^{\prime}, \ Q^{\prime 2}) \cdot d\sigma^{eq}_{inter} )
\end{array}
\end{equation}
\noindent{Prescription D}
\begin{equation}\label{EqPresD}
\begin{array}{lll}
d\sigma^{ep} & = & {\displaystyle\sum_q} ( \int d x^{\prime} f_q (x^{\prime}, (Q^2 + Q^{\prime 2})/2) \cdot d\sigma^{eq}_{lepton}\\
&& + \int d x^{\prime} f_q (x^{\prime},(Q^2 + Q^{\prime 2})/2) \cdot d\sigma^{eq}_{quark}\\
&& + \int d x^{\prime} f_q (x^{\prime},(Q^2 + Q^{\prime 2})/2) \cdot
				d\sigma^{eq}_{inter} )
\end{array}
\end{equation}
\noindent{Prescription E}
\begin{equation}\label{EqPresE}
\begin{array}{lll}
d\sigma^{ep} & = & {\displaystyle\sum_q} ( \int d x^{\prime} f_q (x^{\prime}, \ Q^{\prime 2}) \cdot
d\sigma^{eq}_{lep}\\
& & + \int d x^{\prime} f_q (x^{\prime}, \ Q^2) \cdot d\sigma^{eq}_{quark} \\
& & + \int d x^{\prime} \sqrt{f_q (x^{\prime}, \ Q^2)}
						\sqrt{f_q (x^{\prime}, \ Q^{\prime 2})}
			\cdot d\sigma^{eq}_{inter} )
\end{array}
\end{equation}
\noindent In Eqs. (6-10): the first line collects the expression where the $Z$ boson is emitted from the lepton line ($d\sigma^{ep}_{leptonic}$), the second line production from the quark line ($d\sigma^{ep}_{hadronic}$), and the last line contains the interference of these two production mechanisms ($d\sigma^{ep}_{inter}$).

\noindent At this point we want to introduce a simple and practical proposal to  calculate the process $e + p \to e + Z + X$ by making a modification in the way that we implement the convolution of the PDFs and the amplitude of the quark subprocesses as follows:
\begin{equation}
\begin{array}{lll}
d\sigma^{ep}_{tot} & = & {\displaystyle\sum_q} \int d x^{\prime} \, d\sigma^{eq}_{tot}\\
\end{array}
\end{equation}
where
\begin{equation}
d\sigma^{eq}_{tot}= \frac{2(2\pi^{-5})}{\hat{s}}
              \frac{1}{4} \mid {\cal M}^{eq}_{tot} \mid^{2} d\Gamma_3
\end{equation}
with
\begin{equation}
\mid {\cal M}^{eq}_{tot} \mid^{2} = \mid                                                                                         \sqrt{f_q (x^{\prime}, \ Q^{\prime 2})} \cdot {\cal M}
^{eq}_{l} + \sqrt{f_q (x^{\prime}, \ Q^2)} \cdot {\cal M}
^{eq}_{q}  \mid ^2
\end{equation}
\noindent Instead of the usual Parton Model in which one makes the convolution of the PDFs and the amplitude square of the quark proceses in the following form.

\begin{equation}
\begin{array}{lll}
d\sigma^{ep}_{tot} & = & {\displaystyle\sum_{q}} \int d x^{\prime}
f_q (x^{\prime}, \tilde{Q}^{2}) \cdot
d\sigma^{eq}_{tot}\\
\end{array}
\end{equation}
where
\begin{equation}
d\sigma^{eq}_{tot}= \frac{2(2\pi^{-5})}{\hat{s}}
              \frac{1}{4} \mid {\cal M}^{eq}_{tot} \mid^{2} d\Gamma_3
\end{equation}
and
\begin{equation}
\mid {\cal M}^{eq}_{tot} \mid^{2} = \mid {\cal M}^{eq}_{l} + {\cal M}
^{eq}_{q}  \mid ^2
\end{equation}
We can observe that our proposal reproduces the Prescription E defined in Eq.(10).

\bigskip

\noindent At the lowest order in $\alpha$ only the magnitude of the momentum transfer square of the virtual vector boson exchanged determines whether we are observing a deep inelastic scattering or not. Our $Ansatz$ is motivated by the fact that the exchanged vector boson has to carry enough momentum transfer square to penetrate deep in the proton structure and this can not be guaranteed if this momentum transfer square is taken from the same line where  the $Z$ boson is radiated. In other words, when the $Z$ is emitted from the lepton line we have to take $\tilde{Q}^{2} = Q^{\prime 2} $ and when the $Z$ boson is emitted from the quark line then we have to take $\tilde{Q}^{2} = Q^2$.  For the interference term of the lepton and the quark line mechanisms our proposal means to take the geometric mean of the PDFs associated to each of these two production mechanisms, which contribute to the total amplitude square, $\mid {\cal M}^{eq}_{tot}. \mid^{2}$.

\section{Results}
Now we present the numerical results of our calculations using the Standard Model. We take $M_Z = 91.2$ GeV for the mass of the $Z$ boson and $\sin^2 \theta_W = 0.231$ for the electroweak mixing angle \cite{partdata}. We have included in our computations besides the photon-exchange also the $Z$-exchange diagrams. However, as expected, the dominant contributions to the total cross section come from photon exchange and especially from leptonic initial state $Z$ emission \textit{i.e.} the diagram depicted in Fig.~\ref{fig1}(a)\cite{Bohm:1986my,LlewellynSmith:1977bt}. We give results for the case of unpolarized deep inelastic $ep$-scattering with a total energy
$\sqrt{s}$ in the range $300 \leq \sqrt{s} \leq 1,300$ GeV (We remember here that the expected maximal total energy to be reached at the LHeC is $\sqrt{s}\approx 1,300$ GeV). We take cuts of 4 GeV$^2$, 4 GeV$^2$ and 10 GeV$^2$ on $Q^{2}, \ Q^{\prime 2}$ and the invariant mass $W$, respectively. These values are suited for the PDFs reported by Pumplin {\it et al} \cite{Pumplin,Stump}, we use these CTEQ PDFs provided in a $n_f=5$ active flavors scheme.

\noindent There exist already calculations for the total cross section
$\sigma(ep \to eZX)$, using different PDFs and cuts for the momentum transfer square of the exchanged boson and for the invariant mass $W$ \cite{Bohm:1986my,LlewellynSmith:1977bt,Salati:1982gv,Altarelli:1985jp,Baur:1991pp}

\begin{figure}[ht]
\centerline{\includegraphics[width=6.0in]{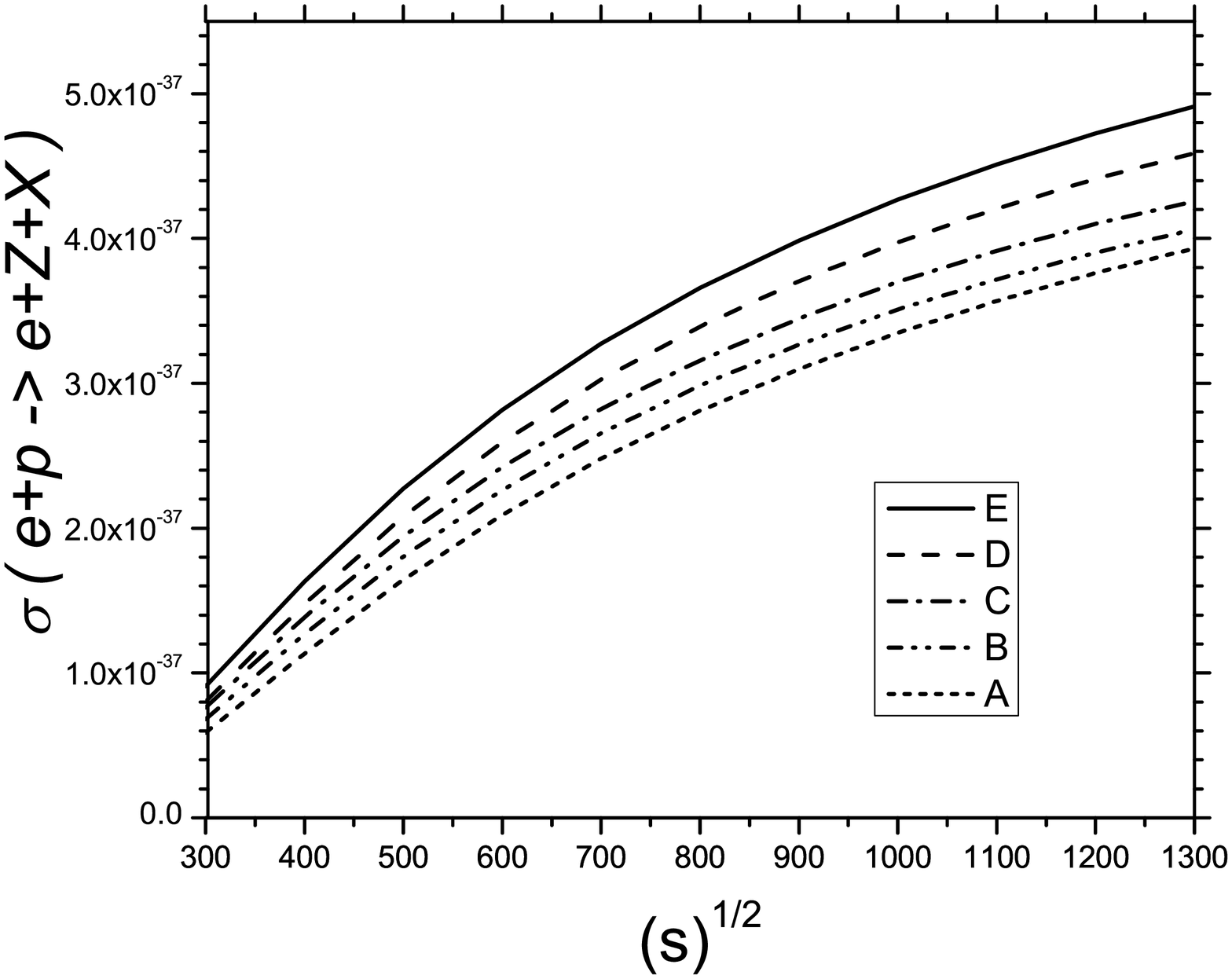}}
\vspace*{8pt}
\caption{Cross section rates for  $Z$-production through the process $e + p \to e + Z + X$ with a total energy $\sqrt{s}$ in the range $300 \leq \sqrt{s} \leq 1,300$ GeV, for the different options that we have taken for doing the convolution of the PDFs with the amplitude of the quark processes, see Eqs.(3.6-3.10).\protect\label{fig2}}
\end{figure}

\noindent Our results for the total cross section by taking the different prescriptions given in Eqs.(6-10), for $\sqrt{s}=300$ GeV (HERA) and 1,300 GeV (LHeC), are showed in Table I. We show in Fig.~\ref{fig2}, the results of the total cross section as a function of the total energy of the $ep$ system $\sqrt{s}$. We can observe in this plot the evolution of the separation of the results of the cross section rates depending on the prescription used. It is clear that differences in the rates of the total cross section can reach up to 25\% depending on the prescription used, for $\sqrt{s} \approx 1,300$ GeV, which is the expected maximal total energy for the system $ep$ to be reached at LHeC.

\noindent We found the rates for  $\sigma(ep \to eZX)$ as given in Table 1. These values yield to a production of $Z$-bosons that we show in Table 2, for the different prescriptions we have chosen for making the convolution of the amplitudes $eq$ subprocesses and the PDFs. We can observe in Table 1, that the differences in the rates of the total cross section can reach up to 25\%  for the different choices of $\tilde{Q}^{2}$ for the expected maximal total energy for the system $ep$ to be reached at LHeC. This difference (25\%) corresponds  to a difference of $\sim 10^4$ produced $Z$ bosons, at the planned LHeC, by taking an integrated luminosity of 0.1 $ab^{-1}/year$ \cite{partdata}

By making very simple calculations, we succeeded in showing the importance of carefully taking the phrase so used in many articles "we identify $\tilde{Q}^{2}$ with the momentum transfer square as usual." And also the phrase "We have proved that our numerical results do not depend strongly on this identification". This kind of phrases were valid some years ago, because the energy and luminosity reached at HERA predicted a production of 40-50 $Z$ bosons per year, then the prediction that the ambiguity of
$\tilde{Q}^{2}$ could give a difference of 11.8\% between prescription C and B (See Table 1) was something that was hardly to be seen ($\sim 5 -6 $). But the same difference (the difference between prescriptions C and B at LHeC is 4.7 \%) at LHeC implies $\sim$ 470 $Z$ bosons.

\section{Conclusions}
In this letter we discuss $Z$-production at the planned LHeC at CERN, where the expected center of mass energy of $\sqrt{s}\approx 1,300$ GeV is 4 times larger than the maximal total energy reached at HERA, $\sqrt{s}\approx 300$ GeV. The luminosity planned to be reached at LHeC is at least two orders of magnitude larger than luminosity reached at HERA. We have presented the calculation of the cross section of the $Z$-production in deep inelastic electron-proton scattering in a range of energies which is expected to be available in the in near future, in the framework of the Standard Model, by using the Parton Model. We showed how strong is the dependence of the cross section rates on the scale energy parameter $\tilde{Q}^{2}$.

\noindent In this note we have analyzed at tree level the production of Z bosons in deep inelastic $ep$-scattering, finding a strong dependence ($\sim$ 25\%) on the $\tilde{Q}^{2}$-prescription adopted in the PDFs. We want to make clear that our aim in this letter  is to point out the necessity of a detailed revision of the Parton Model in order to use the results provided by the LHeC and even the results obtained by the  LHC to look for Physics beyond the Standard Model. Our paper can be seen as a first step to look for a possible unique identification of $\tilde{Q}^{2}$.

We do not doubt that the PDFs will be obtained with a great precision and that $\tilde{Q}^{2}$ should be identified with the momentum transfer square. Our aim in this letter is to point out that in the case of the production of $Z$ boson or any other particle, we have two transfer square momenta: ${Q}^{2}$ and $Q^{\prime 2}$. Also, we have demonstrated that although this fact was not important at HERA energies, now will be very important at LHeC energies. A detailed comparison of the phenomenological predictions and the experiment results could elucidate on the correct prescription in order to do the convolution of the PDFs and  the amplitudes of the quark processes which leads to the calculation of the $e + p \to e + Z + X $.

We end this letter with the following comment. We can use the dependence of the cross section as a function of the dimensionless variables $y$ or as a function of dimensionless variables $x$ and $y$ in order to use the energy of the final electron and the angle that the momentum of the outgoing electron ($p'$) makes with the momentum of the incident electron (p) to define kinematical regions in which the leptonic or the hadronic contribution to the total cross section predominates   \cite{Bohm:1986my}. In such case, there will be only one important momentum transfer square, $Q^{\prime 2}$ or ${Q}^{2}$. Hence, in such regions there will not more problem with the identification of the scale energy parameter $\tilde{Q}^{2}$.

\bigskip

\begin{table}[h]
\noindent{Table\,1. Cross section rates in $10^{-37} cm^2$ for $Z$ production for $\sqrt{s}=$ 300 GeV (HERA) and 1,300 GeV (LHeC), for the different prescriptions that we have taken for making the convolution of the PDFs with the amplitude of the quark processes.}

\vspace{-0.5cm}

\begin{center}
\begin{tabular}{|c|c|c|c|c|c|}
\hline
 &$\tilde{Q}^{2}=M^2_Z$ & $\tilde{Q}^{2}=Q^2$ &$\tilde{Q}^{2}=Q^{\prime 2}$& $\tilde{Q}^{2}=(Q^2 + Q^{\prime 2})/2$ & Our Proposal\\
\hline
$ \sqrt{s}$ (GeV) & $\sigma_A$ & $\sigma_B$ &$\sigma_C$ & $\sigma_D$ & $\sigma_E$\\
\hline
300 & 0.587 & 0.682 & 0.763 & 0.803 & 0.909 \\
\hline
1,300 & 3.934 & 4.064 & 4.257 & 4.590 & 4.912 \\
\hline
\end{tabular}\label{ta1}
\end{center}
\end{table}

\bigskip

\begin{table}[h]
\noindent{Table\,2. Number of $Z$ bosons that will be produced through $e + p \to e + Z + X$ taking $\sqrt{s}=1,300$ GeV (LHeC) and taking an integrated luminosity of 0.1 $ab^{-1}/year$, for the different prescriptions that we have adopted for making the convolution of the PDFs with the amplitude of the quark processes.}

\vspace{-0.5cm}

\begin{center}
\begin{tabular}{|c|c|c|c|c|c|}
\hline
 &$\tilde{Q}^{2}=M^2_Z$ & $\tilde{Q}^{2}=Q^2$ &$\tilde{Q}^{2}=Q^{\prime 2}$& $\tilde{Q}^{2}=(Q^2 + Q^{\prime 2})/2$ & Our Proposal\\
\hline
$\, \sqrt{s}$ (GeV)\, & $N_A$ & $N_B $ &$N_C $ & $N_D$ & $N_E$\\
\hline
1,300 & $3.93\times 10^4$ & $4.06\times 10^4$  & $4.26\times 10^4$ & $4.59\times 10^4$  & $4.91\times 10^4$  \\
\hline
\end{tabular}\label{ta2}
\end{center}
\end{table}

\bigskip

\section*{Acknowledgments}

This work was supported in part by the {\it Consejo Nacional de Ciencia y
Tecnolog\'{\i}a} (CONACyT) and {\it Sistema Nacional de Investigadores (SNI) de
M\'exico}.

\end{document}